\documentclass[twocolumn,showpacs,preprintnumbers,amsmath,amssymb]{revtex4}

\usepackage{dcolumn}
\usepackage{bm}
\usepackage{graphicx}
\graphicspath{{./}}

\begin{document}

\preprint{APS/123-QED}

\newcommand{\beq}{\begin{equation}}
\newcommand{\eeq}{\end{equation}}
\newcommand{\beqa}{\begin{eqnarray}}
\newcommand{\eeqa}{\end{eqnarray}}
\newcommand{\noid}{\noindent}
\newcommand{\ssones}{1_{s \times s}}
\newcommand{\opf}{\frac{1}{4}}
\newcommand{\ma}{\mathbb A}
\newcommand{\mt}{\mathbb T}
\newcommand{\msqp}{\mathbb SQP}
\newcommand{\mhpp}{\mathbb HPP}
\newcommand{\mrw}{\mathbb RW}
\newcommand{\etal}{{\it et al}.~}

\title{Fractal geometry of critical Potts clusters}

\author{J.~Asikainen$^1$, A.~Aharony$^2$, B.~B.~Mandelbrot$^3$,
E.~M.~Rauch$^3$ and J.~-P.~Hovi$^{2,3}$}

\affiliation{
$^1$ Helsinki Institute of Physics and
Laboratory of Physics, Helsinki University of
Technology \\
$^2$ Raymond and Beverly Sackler Faculty of Exact Sciences,
School of Physics and Astronomy Tel Aviv University, Ramat Aviv
69978, Tel Aviv, Israel \\
$^3$ Artificial Intelligence Laboratory, Massachusetts
Institute of Technology, Cambridge, MA 02139 USA
}

\date{\today}

\begin{abstract}
Numerical simulations on the total mass, the numbers of bonds on
the hull, external perimeter, singly connected bonds and gates
into large fjords of the Fortuin-Kasteleyn clusters for
two-dimensional $q$-state Potts models at criticality are
presented. The data are found consistent with the recently derived
corrections-to-scaling theory. However, the approach to the
asymptotic region is slow, and the present range of the data does
not  allow a unique identification of the exact correction
exponents.
\end{abstract}

\bigskip

\maketitle

\section{Introduction}
\label{intro.sect}

$q$-state Potts models have played an important role in
condensed matter physics\cite{Wu82}.
Here we study geometrical
aspects of the critical Potts clusters, in two dimensions.
The $q$-state Potts model \cite{Pot52} is defined
through the Hamiltonian
\beq
{\cal H} = - K \sum_{\langle i,j \rangle} (\delta_{\sigma_i,\sigma_j} - 1),
\label{hamiltonian.eq}
\eeq
where $\langle i,j \rangle$ denotes the summation over nearest
neighbor sites $i,j$, the spin variable $\sigma_i$ can take any of
the values $1,2,\ldots,q$ and $K$ is the thermal coupling with the
factor $1/k_{\rm B} T$ absorbed in it. It is possible to extend
$q$ to real values \cite{Wu82}, but here we concentrate on the
Potts model with integer values of $q$. In particular, we study
Potts models with $q=1,2,3,~{\rm and}~4$, where the thermal phase
transition at the critical inverse temperature $K_c = \ln{(1 +
\sqrt{q})}$~\cite{Wu82} is of second order. Below we present
simulations at $K=K_c$.

One defines the fractal clusters in the Potts model through the
Fortuin-Kasteleyn (FK) \cite{For72} cluster decomposition, which
states that the model can be mapped onto a general percolation
model.
%
%
%
%
%
The partition function of the Potts model ${\cal Z} = {\rm
Tr}_{\sigma} e^{\cal H}$ can be expressed in terms of bond
variables as ${\cal Z} = {\rm Tr}_{\rm bonds} p^b (1-p)^n
q^{N_c}$, where $b$ is the number of bonds and $n$ is the number
of interactions that did not form a bond in a configuration with
$N_c$ clusters \cite{Swe87}.
%
%
%
%
%
%
Here, $p=1-e^{-K}$, and ${\rm Tr}_{\rm
bonds}$ means a summation over bonds.
Thus, the problem of a thermal lattice model can be mapped to a
graph problem. The FK decomposition has been the starting point
for efficient cluster algorithms \cite{Swe87,Wol89} for simulation
of spin models.

The Potts model has been shown to exhibit a rich critical
behavior, and it has been related to a number of problems in
lattice statistics \cite{Wu82}. Although of great theoretical
interest in itself, it also has many experimental realizations.
The $1$-state Potts model is equivalent to a bond percolation
problem \cite{Wu82}, and the $2$-state Potts model is the same as
the Ising model \cite{Ising}. The $q=3$ Potts model has been shown
to describe absorbed monolayers on two-dimensional (2D) lattices
\cite{Ale75,Bre77}. Domany {\it et al.} \cite{Dom77} suggested
that $N_2$ absorbed on krypton-plated graphite should exhibit the
same critical behavior as the $q=4$ Potts model. More references
of the experimental realizations can be found in the review
article by Wu \cite{Wu82}.

Here we study geometrical aspects of the critical Potts clusters
in two dimensions. This is in direct analogy with the geometry of
percolation clusters, which has been widely studied
\cite{Sta95,Sal87,Aiz99,Gro86}. Specifically, we measure the
fractal dimensions $D_M,~D_H,~D_{EP},D_{SC},~{\rm{and}}~D_{G}$
describing the scaling of the cluster's mass, hull, external
accessible perimeter \cite{Gro86}, singly connected bonds
\cite{Con82} and the gates to {\it narrow-gate} fjords
\cite{Aiz99}, respectively, with its radius of gyration $R$. As
emphasized by Coniglio \cite{Con89}, many of these fractal
dimensions have been derived analytically \cite{Sal87}. Some
others have been found more recently \cite{Aiz99,Dup00}. Although
there exists much numerical work on the percolation clusters (i.e.
$q=1$), we are not aware of
any detailed numerical study of most
of the above mentioned quantities for $q>1$, and especially when
$q$ approaches the critical value $q_c = 4$.

Section \ref{simulation.sect} describes the numerical methods used
in the simulation of the Potts models. Our numerical simulations
show that the asymptotic power law dependence of the various
masses on $R$ is approached relatively slowly, and therefore the
analysis of the data must include {\it correction terms},
particularly as $q$ approaches $q_c$. The theory developed to
obtain these correction terms \cite{Aha02} is briefly summarized
in Section \ref{corrections.sect}. We compare our numerical data
with the exact predictions in Section \ref{results.sect}. Finally,
we present the summary and conclusions in Section
\ref{conclusions.sect}.

\section{Simulation}
\label{simulation.sect}

Numerical simulations of spin models have developed from the local
spin flip type algorithm \cite{Met53} to the more advanced cluster
algorithms \cite{Swe87,Wol89}. Our simulations were done on a 2D
square lattice with both open and periodic boundary conditions.
Clusters of the $q$-state Potts model were generated using the
Swendsen-Wang algorithm~\cite{Swe87}, which is based on the
cluster decomposition by Fortuin and Kasteleyn~\cite{For72}. The
size of the system in our simulations was $4096^2$ spins for all
$q$. Figure \ref{fig1} shows sample clusters for different values
of $q$.

All simulations were started with a homogeneous initial condition,
with all spins initially parallel to each other. First we
thermalized the system to allow the model to
equilibrate.
Thermalization was checked by measuring both the energy per spin
$e$, directly from the Potts Hamiltonian, and the magnetization
per spin $m$, using the representation of Potts spins in a $q-1$
dimensional space \cite{Wu82}.

Thermalization of large spin systems takes a very long time. The
quantities of interest in this work show a relatively slow
approach to the asymptotic values. Thus, extremely large lattices
are required for the analysis of the scaling behavior of the
cluster subset masses. When performing simulations on lattices of
linear size $L = 2^{12} = 4096$, about $20000$ Monte Carlo steps
(lattice sweeps) are needed to equilibrate the system.

We devised a simple method to overcome the problem with long
thermalization times. We started the thermalization with a small
lattice of size $L_1$ and thermalized it. We then periodically
copied the spin configuration of the small lattice to a lattice
with twice as large a size, $L_2 = 2 \times L_1$, and thermalized
it. We continued this process until the desired system size was
reached. In practice, it is recommended to compare the values of
$e$ and $m$ obtained this way with the values obtained from
conventional thermalization to be sure that the system is really
thermalized. Alternatively, one can continue running the
simulation and collect the values of thermodynamical variables as
a function of time and check that there is no increasing or
decreasing trend in them. The thermalization method described
above allows a speed-up by an order of magnitude in thermalization
for a Potts spin system of $1024^2$ spins.

After the spin system was thermalized, we took samples of the
cluster configurations after every $20$ spin updates
(corresponding roughly to the correlation time for the present
system sizes) to get uncorrelated samples. Each cluster of the
present configuration was taken separately under investigation.
However, as a precaution to avoid some of the finite size effects,
we collected the data only from those clusters which did not
involve spins on the boundary. For the remaining clusters we
determined the masses of the cluster subsets. The total mass of
the cluster is defined as the number of occupied bonds in the
cluster. The number of bonds belonging to the hull, external
perimeter and the number of singly connected bonds were counted
using directed walkers that walk on the appropriate cluster
perimeter \cite{Sta95,Gro86}.

To measure the number of gates to fjords of different gate sizes
($S_G$) a walk was initiated on the EP. The walker starts from the
left vacant neighbor of the lower-left site belonging to the
cluster and goes around the cluster on the EP always trying to
turn to the right. At each site, we look at the neighbors on the
left hand side and check whether the sites within a predefined
distance $S_G$ belong to the EP or not. If the sites (and bonds)
up to the distance $S_G$ are not EP sites and the site at the
distance $S_G$ belongs to the EP, the walker is about to enter a
fjord with a gate size $S_G$. All fjords with different gate sizes
were counted during a single walk around the cluster with an array
of boolean variables that indicate whether the walker is within a
fjord of given size. If the walker was in a fjord of a gate size
$S_G$, fjords with gate size $S_G' > S_G$ were not allowed in the
statistics. In practice, all measured values of $S_G$ gave similar
results, and we report only the results for  $S_G = 1$.

As noted by Aizenman \etal \cite{Aiz99}, the scaling concerns
fjords whose size $L_F$ is comparable to that of the cluster, $R$,
and whose gate's width is much smaller than $L_F$. The reason for
this is easy to understand, since small kinks and pits are a
natural part of the fractal cluster's perimeter. Thus, only large
enough fjords were included in the statistics. This was taken into
account by choosing a suitable parameter $s$, and counting only
fjords with $L_F > sR$.

Since any thermalized spin configuration contains clusters of many
sizes, we collected the data in multiplicatively increasing bins
of the size of the cluster. Each bin contained the clusters of
sizes within $[R_i,R_{i+1} = \sqrt{2}~R_i]$.

\section{Corrections to scaling}
\label{corrections.sect}

In our recent publication \cite{Aha02}, we derived theoretically
the {\it corrections-to-scaling} terms for the various cluster
subset masses. The corrections arise from three different sources:
(a) using the renormalization group approach, (b) mapping to the
coulomb gas, and (c) considering the uncertainty of the correct
measure for the linear cluster size, which implies corrections of
the order of $1/R$.

The first correction relates to the {\it dilution} field $\psi$,
which is generated under renormalization even when one starts with
the non-diluted case \cite{Car80}. Solving the renormalization
group recursion relations (RGRR's) for $\psi(\ell)$, where $\ell$
is the scale variable, substituting the solution to the RGRR's for
the field $h_S$ conjungate to the density $\rho_S = M_S/R^d$ and
finally calculating the scaling of the density $\rho_S(\ell)$ with
the cluster's linear size $R=e^\ell$ yields the following
predictions for the approach of each mass $M_S$ to the
asymptotics. Here, $S=M,~H,~EP,~SC$ or $G$.

In the $q=4$ case, the renormalization group calculation is exact,
yielding logarithmic corrections to the scaling of $M_S(R)$:
\beqa
M_S &\propto& R^{D_S}(\log R + B \log(\log R)+ E)^{-c_S/a} \nonumber\\
    && \times (1+{\cal O}(\log\log R/\log R)),
\label{logcorr.eq} \eeqa with $D_S=y_S(q=4)$ and $c_S/a$ as given
in Table \ref{theory.tab} (see below) . Note that $B$ is
universal, and the non-universal constant $E$ is the same for all
$S$. Equation (\ref{logcorr.eq}) generalizes the logarithmic
corrections of Cardy \etal \cite{Car80}.

We were recently directed to a previous study by
Vanderzande and Marko \cite{Van93}, in which they considered
corrections to scaling for percolative properties of the 
$q=4$ Potts model. They also found logarithmic corrections
to scaling in aggreement with those presented here.

For $q<4$, to leading order in $\epsilon' = \sqrt{4-q}$, the same procedure
yields
\begin{equation}
M_S \propto R^{D_S}(1-\hat B R^{-\theta})^{-c_S/a} \approx
R^{D_S}(1+f_S R^{-\theta}),
\label{powcorr.eq}
\end{equation}
where $D_S \approx y_S-c_S \epsilon'$ and $\theta \approx 2a
\epsilon'$. Note that to the lowest order in $\epsilon'$, the
ratios $f_S/f_{S'}$ are universal, being equal to $c_S/c_{S'}$.
This is similar to analogous ratios for thermodynamic properties
in the usual $\epsilon$-expansion \cite{Aha80}. This universality
should hold to all orders in $\epsilon'$. Expanding the exact
$D_S$ (Table
\ref{theory.tab}) in $\epsilon'$ yields $c_S$. Using also
$a=1/\pi$ \cite{Car80,Aha02} yields our predictions for $c_S/a$
(given in Table
\ref{theory.tab}), to be used in the fitting procedure.

The second source of corrections involves new contributions to the
relevant pair correlation functions in the Coulomb gas
representations. den Nijs \cite{Nij83} derived such corrections to
the order parameter correlation function. Since correlation
exponents $x$ are related to the fractal dimension via $D = d -
x$, the correction exponents can be related to the corresponding
correction terms for the scaling of the mass $M_M$, yielding the
leading correction \beq \theta' = 4/g, \label{cgmm.eq} \eeq where
$g$ is the ($q$-dependent) Coulomb Gas coupling constant (see
Table \ref{theory.tab}).

Using a similar approach we found in the case of the hull and the
singly connected bonds that the leading correction exponent is
given by \beq \theta'' = 2/g. \label{cgmhmsc.eq} \eeq We argued
that this correction would also hold for the external perimeter
\cite{Aha02}.

The last source of corrections involves `analytic'
terms, coming e. g. from linear cuts with dimensions $(D_S-1)$,
\cite{partD} or from replacing $R$ by $(R+A)$, since there are
many possible candidates for the correct linear measure of the
cluster. These would imply corrections of relative size $1/R$.

\section{Results}
\label{results.sect}

We now present the numerical data from large scale Monte Carlo
simulations of the Potts models. Our aim is to confirm the exact
predictions of the fractal dimensions $D_S$ in the cases where
they are available and to give numerical estimates for the
exponents that have not yet been calculated exactly. In addition,
we want to numerically confirm the corrections-to-scaling theory
presented in the previous section.

We obtain good agreement with the theoretically predicted values
for most of the fractal dimensions $D_S$. The worst agreement is
found for the exponent of the external perimeter $D_{EP}$ for
$q>2$ Potts models. The reasons for this will be discussed below.
However, fixing the correction terms and performing fits only to
the amplitudes and to the fractal dimensions $D_S$ in the
logarithmic derivatives of Eqs. (\ref{logcorr.eq}) and
(\ref{powcorr.eq}), yielded estimates for the subset fractal
dimensions that agree to the precision of $0.05$ or better with
the theoretical predictions of Table \ref{theory.tab}.

We are not able to give a quantitative numerical proof of the
values predicted for the correction terms. Qualitatively, our
numerically evaluated functions $M_S(R)$ display a complex
corrections-to-scaling behavior that requires more than one
correction term. The theoretical predictions can be fitted
reasonably well to the data keeping the important quantities such
as $\theta$, $\theta'$ (or $\theta''$) and $c_S/a$ fixed while
leaving the amplitudes of individual correction terms free.

We start this section by studying the fractal geometry of clusters
in the site percolation model, which is computationally easier to
simulate. The reasons behind the difficulties in the comparison of
the numerical data with the analytical predictions are discussed.
We then proceed to present our numerical data on the Potts
clusters. In all the figures of this Section, whenever the error
bars are not shown they are smaller than the size of the symbols.

\subsection{Site Percolation Clusters}

For $q=1$ we simulated site percolation on a square lattice of
size $24576^2$, using the Newman-Ziff cluster labeling
\cite{ZN_Alg} which is an improved version of the Hoshen-Kopelman
algorithm \cite{Hos76}. Thus the linear lattice size was $6$ times
larger than in the simulation of Potts clusters with $q>1$.

To get a feeling of what kind of problems are present in the
fitting procedure when the theoretically predicted correction
terms of Eqs. (\ref{powcorr.eq}) and (\ref{logcorr.eq}) are fitted
to the numerical data on the cluster subset masses $M_S(R)$, let
us consider as an example the scaling of the number of singly
connected bonds. Figure \ref{fig2} illustrates the scaling of
$M_{SC}(R)$ with the cluster size $R$ on a double logarithmic
scale. The solid line in the main figure indicates the predicted
slope. The fit to the data yields an estimate for the asymptotic
fractal dimension $D_S$ which is less than $0.01$ off the exactly
known value $D_{SC}=3/4$.

Although the asymptotic scaling regime $M_{SC}(R) \propto
R^{D_{SC}}$ can be seen here, there are difficulties in the
extraction of the correction terms of Eq. (\ref{powcorr.eq}). The
%
%
smallest value of $R$ included in the linear fit to the data on $\log -
\log$ scale in Fig. \ref{fig2} corresponds to the regime where the
influence of the correction terms are about to vanish, thus justifying
fitting without any correction terms. The saturation to the asymptotics
can be seen more clearly in the inset of Fig. \ref{fig2} where data are
scaled with the predicted asymptotic behavior $M_{SC}(R)/R^{D_{SC}}$.
The inset shows that at about $R \approx 300$ the correction terms can be
neglected in this case.
%
%
However, at the same point the statistics becomes so
noisy, making a precise estimation of the correction terms
difficult. An additional difficulty arises from the fact that the
finite size of the lattice is not taken into account in any way in
the finite size scaling form of Eq. (\ref{powcorr.eq}). Due to the
finite system sizes, statistics of the large cluster is biased in
such a way that only the compact clusters fit in the lattice
without touching  the boundaries. The extended clusters having for
example more EP sites than compact clusters with the same radius
of gyration $R$, are absent. This bias cannot be taken into
account by any known correction terms. We tried to extrapolate the
data from different system sizes to obtain an asymptotic curve for
an infinitely large system, but the statistics is far from
sufficient for such a procedure.

\subsection{Potts Clusters}

In the case of Potts clusters, system sizes that can be
used in the simulations are much smaller
than those
in the site percolation case, since in addition to the
spin variables, also bonds must be stored in the computer
memory. This causes the finite size effects to be even more
pronounced than those present in the site percolation model simulations.
In addition, the correction exponents in the $1<q<4$ Potts models are smaller
than those in the $q=1$ case. Also, the logarithmic corrections
present in the $4$-state Potts model are weaker than any of the
power law corrections in $q<4$ models.
Thus, the influence of the corrections-to-scaling terms
extends to much larger values of $R$.

The data analysis was done by fitting the theoretical predictions
of Sect. \ref{corrections.sect} to the data. The nonlinear fitting
was done using the Levenberg-Marquardt method \cite{NR}. The
measure for the quality of fits is $\chi^2$. Values of $\chi^2$
close to one indicate a good fit. We determined the error bars of
the fractal dimensions by fixing $D_S$ to a range of values around
the theoretically predicted one, and performing a fit to the
amplitudes for each such value. The error bar on $D_S$ was fixed
as the range within which $\chi^2$ remained smaller than $2$. One
can perform the fit directly to $M_S(R)$, or to the effective
fractal dimensions $D_S^{\rm eff}(R)$ (which is the logarithmic
derivative of $M_S(R)$ \cite{Sta95}).

We found that for $q<4$, fitting directly to the mass worked
out better. If one wants to fit to the measured data directly, it
is recommended to divide the measured data on the cluster subset
masses by the exactly known asymptotic behavior to avoid problems
with numerical accuracy. Thus, for $q<4$, the fits are done to
the form
\begin{equation}
M_S / R^{D_S} = E_S (1 + f_S R^{-\theta} + f_S' R^{-\theta'} + g_{S}/R),
\label{powfit.eq}
\end{equation}
where
$E_S,f_S,f_S',g_S,\theta$ and $\theta'$ are the fitting parameters,
and the fractal dimensions $D_S$ are kept fixed.

In the $q=4$ case, on the other hand, we found the fitting
to the effective exponent $D_S^{\rm eff}$ better, and the fits
were thus done to
\begin{eqnarray}
D_S^{\rm eff} &=& d\log M_S/d\log R \nonumber \\
&\approx& D_S - \frac{(c_S/a) (B+C+\log R)} {(C+\log R)(E+\log R+
B \log(C+\log R))}
\nonumber \\
&& + Z/\log R.
\label{logfit.eq}
\end{eqnarray}
Note that we have
replaced the $\log\log R$ term in Eq. (\ref{logcorr.eq}) by the
more general $\log(C+\log R)$.
Also, the logarithmic derivative of the higher order term on the
RHS of Eq. (\ref{logcorr.eq}) was approximated by a simpler form
$1/\log R$. In the fitting procedure, the possibility of having
many candidates for the correct linear measure of the cluster size
was taken into account by allowing $R$ to adjust to $R+A$ (see
Sect. \ref{corrections.sect}).

Below, the results of our numerics are summarized for the various
subsets. We do not go into the details of numerical estimates for
the various amplitudes in the nonlinear fits, since due to the
relatively small range of the data and many fitting parameters the
estimated error bars are large, and allow no comparison with, for
example, the predicted amplitude ratios. Also, precise estimation
of the correction exponents $\theta$, $\theta'$
(or $\theta''$) as well as the parameters $c_S/a$ is impossible
with the presently available range of data. Instead, we keep the
correction exponents
(or parameters)
fixed and try to extrapolate the fractal dimensions $D_S$, and to
demonstrate that the predicted forms of scaling in Eqs.
(\ref{logfit.eq}) and (\ref{powfit.eq}) are consistent with our
numerical data. Specifically, in all the fits for $1 \le q < 4$,
$\theta$ and $\theta'$ (or
$\theta''$) of Eq. (\ref{powfit.eq})
were kept fixed (at the predicted values), while the amplitudes of
each correction term were allowed to adjust. In the $q=4$ case,
only $c_S/a$ was fixed in Eq. (\ref{logfit.eq}).

The logarithmic corrections are most important for the singly
connected bonds at $q=4$, where theory predicts that $D_{SC}=0$
(see Table \ref{theory.tab}). Indeed,  the solid line in Fig.
\ref{fig3} shows that a fit to Eq. \eqref{logfit.eq} is consistent
with this theoretical prediction. In contrast, a fit with a single
power law correction term $\theta=1/2$ (dashed line in Fig.
\ref{fig3}) extrapolates to a wrong value near $D_{SC}=0.21$!

\subsubsection{Mass}
Fig. \ref{fig4} shows an example of the fit to the curve
$M_{M}(q=3)/R^{D_M}$. In the fitting procedure, $D_M$, $\theta$
and $\theta'$ were kept fixed while $E_M,~f_M,f'_{M}$ and $g_M$
were allowed to fit. The value $\chi^2=1.17$ indicates that Eq.
(\ref{powfit.eq}) gives a good representation of the data. Our
numerical estimates for the fractal dimensions $D_M(q)$,
determined as the range of values for which one has $\chi^2<2$,
are $1.90 \pm 0.01$, $1.87 \pm 0.01$, $1.85 \pm 0.02$ and $2.05
\pm 0.15$ for $q=1,2,3,4$, respectively. These are in good
agreement with the theoretical predictions.

\subsubsection{Hull}
Figure \ref{fig5} shows a fit to the number of the bonds belonging
to the hull in the $q=4$ Potts model. In this particular fit,
$D_H$ and $c_H/a$ were kept fixed while $B,~C,~Z$ and
$A$ were free to adjust. The value of $\chi^2=1.11$ indicates that
Eq. (\ref{logfit.eq}) fits the data well.  Our numerical estimates
for the fractal dimensions $D_H(q)$ are $1.75 \pm 0.01$, $1.66 \pm
0.01$, $1.59 \pm 0.03$ and $1.50 \pm 0.01$, for $q=1,2,3,4$,
respectively. Agreement with the theoretical predictions is
excellent as can be seen by comparison with the values in Table
\ref{theory.tab}.

\subsubsection{External Perimeter}

Figure \ref{fig6} shows an example of the
fit to the external perimeter data in the $q=2$ Potts model. The
fractal dimension $D_{EP}$ and the correction exponents $\theta$
and $\theta''$ were kept fixed at the predicted values while
$E_{EP},~f_{EP},f'_{EP}$ and $g_{EP}$ were free to adjust yielding
$\chi^2=1.77$, implying a reasonably good agreement with Eq.
(\ref{powfit.eq}). Again, in the fits for $q<4$, $\theta$ and
$\theta''$ of Eq. (\ref{powfit.eq}) were kept fixed and in the
$q=4$ case, $c_{EP}/a$ was fixed. The numerical estimates $1.33
\pm 0.05$, $1.36 \pm 0.02$, $1.40 \pm 0.02$ and $1.48 \pm 0.02$
for $q=1,2,3,4$, respectively, agree with the exact predictions.

\subsubsection{Singly connected bonds}
In Fig. \ref{fig7}, we show the number of singly connected bonds
$M_{SC}(R)/R^{D_{SC}}$ against the cluster size $R$ in the $q=2$
Potts model. The fractal dimension $D_{SC}$ and the correction
exponents $\theta$ and $\theta''$ were kept fixed to the predicted
values while $E_{SC},~f_{SC},f'_{SC}$ and $g_{SC}$ were allowed to
fit. The value $\chi^2=1.21$ implies good agreement with Eq.
(\ref{powfit.eq}).
 The numerical estimates for the fractal
dimensions are $0.75 \pm 0.02$, $0.55 \pm 0.03$, $0.35 \pm 0.07$,
for $q=1,2,3,4$, respectively. All the estimates for the fractal
dimensions $D_{SC}$ are in good agreement with the theoretical
predictions. However, the large value of $\chi^2 \approx 3$ in the
$q=4$ case indicates some discrepancy between Eq.
(\ref{logfit.eq}) and the data.

\subsubsection{Gates to Fjords}
Figure \ref{fig8} shows our numerical data for the number of gates
to narrow-gate fjords. The figure shows fits to the data along
with the estimates for the fractal dimensions $D_G$. Our estimate
$D_G(q=1) = -0.9 \pm .05$ agrees with the exact prediction $D_G =
-11/12 \approx -0.92$ \cite{Aiz99}. Here, only a linear fit to the
data on the double logarithmic scale was considered, since the
scaling regime for the presently available cluster sizes is rather
narrow. The parameter $s$ governing the minimal ratio of the fjord
size to the cluster size that we used was in the range $0.1 \le s
\le 0.2$. The actual choice for the value of $s$ does not affect
the scaling law, but it merely determines the range where the
power law behavior $M_G \sim R^{D_G}$ starts (decreasing $s$
shifts the maximum of the curves $M_G(R)$ to the left). Our
numerical estimates for $|D_G(q)|$ decrease with increasing $q$.
Our estimates for $D_G(q)$ are $-0.90 \pm 0.05$, $-0.71 \pm 0.05$,
$-0.63 \pm 0.05$ and $-0.59 \pm 0.05$ for $q=1,2,3,4$,
respectively. Our numerical estimates together with the
theoretical predictions for all fractal dimensions $D_S$ are
summarized in Table \ref{numerics.tab}.

\section{Conclusions}
\label{conclusions.sect}

The present paper examined the fractal geometry of the Potts
clusters at the critical temperature. The aim was to find
numerical evidence on the exactly derived subset fractal
dimensions $D_S$ \cite{Sal87,Con89,Aiz99,Dup00} and to give
estimates on the dimensions for which there is no exact
prediction. We gave the first numerical estimate of the negative
fractal dimensions $D_G$, describing the scaling of the gates to
fjords \cite{Aiz99}.

Analysis of our numerical data revealed a slow and complex
approach to the asymptotic behavior. If this is neglected in data
analysis, wrong numerical estimates for the dimensions $D_S$
follow. Using the corrections-to-scaling terms derived in our
earlier publication \cite{Aha02} in the fitting procedure,
excellent agreement with most of the exact dimensions and data was
found. The present quality and range of data does not allow a
unique quantitative confirmation of the exact correction
parameters.

To summarize, in the comparison between theory and numerics, extreme
caution is needed in the extraction of the fractal dimensions
$D_{S}$ from the numerical data. The corrections-to-scaling
theory presented already implies that
the finite size effects arising from the finite cluster size
are strong. In addition, effects coming from the finite
lattice size lead to an uncontrollable bias that is very
difficult to handle.

\section{Acknowledgments}
\label{acknowledgments.sect}

We would like to thank Carlo Vanderzande for
drawing our attention to their earlier work.
We acknowledge support from the German-Israeli Foundation. This
work has also been supported in part by the Academy of Finland
through its Center of Excellence program. J. A. wishes to thank
the Vaisala Foundation for financial support.

\bibliographystyle{prsty}
%


\begin{table}[h]
\centering
\begin{minipage}{8.5 cm}
\renewcommand{\arraystretch}{1.2}
\vspace{.4cm}
{\begin{tabular}{c c c c c c c}
\hline
\hline
 &$D_S$ & $q=1$ & $q=2$ & $q=3$ & $q=4$ & $c_S/a$ \\
\hline $g$ & & $\frac{8}{3}$&3&$\frac{10}{3
}$& 4 & \\
\hline $M$ & $(g+2)(g+6)/(8g)$ & $\frac{91}{48}$&$\frac{15}{8}$
&$\frac{28}{15}$&$\frac{15}{8}$&$\frac{1}{16}$ \\
$H$ & $1+2/g$& $\frac{7}{4}$&$\frac{5}{3}$&$\frac{8}{5}$&$\frac{3}{2}$&
$-\frac{1}{4}$\\
${EP}$& $1+g/8$ & $\frac{4}{3}$&$\frac{11}{8}$&$\frac{17}{12}$&
$\frac{3}{2}$& $\frac{1}{4}$\\
${SC}$ & $(3g+4)(4-g)/(8g)$& $\frac{3}{4}$&$\frac{13}{24}$&
$\frac{7}{20}$&$0$& $-1$\\
\hline
$\theta$ & $4(4-g)/g$ & 2 & $\frac{4}{3}$ & $\frac{4}{5}$ & 0 (log) & \\
$\theta'$ & $4/g$ & $\frac{3}{2}$ & $\frac{4}{3}$&$\frac{6}{5}$&1& \\
$\theta''$ & $2/g$ & $\frac{3}{4}$ & $\frac{2}{3}$&$\frac{3 }{
5}$&$\frac{1}{2}$ &
 \\
\hline
\hline
\end{tabular}}
\caption{Exact theoretical predictions.}
\label{theory.tab}
\end{minipage}
\end{table}

\begin{table}[!ht]
\begin{center}
\renewcommand{\arraystretch}{1.2}
\vspace{.4cm}
\begin{tabular}{c c c c c c c c c}
\hline
\hline
$S$ & \multicolumn{2}{@{}c}{$D_S(q=1)$} & \multicolumn{2}{@{}c}{$D_S(q=2)$} & \multicolumn{2}{@{}c}{$D_S(q=3)$} & \multicolumn{2}{@{}c}{$D_S(q=4)$} \\
\hline
& n & e & n & e & n  & e  & n & e \\
\hline
$M$ & $1.90 (1)$ &  $\frac{91}{48}$ & $1.87 (1)$ & $\frac{15}{8}$ & $1.85 (2)$ &$\frac{28}{15}$ & $2.05 (15)$ & $\frac{15}{8}$ \\

$H$ & $1.75 (1)$ & $\frac{7}{4}$ & $1.66 (1)$ & $\frac{5}{3}$ & $1.59 (3)$ & $\frac{8}{5}$ & $1.50 (1)$ & $\frac{3}{2}$ \\

${EP}$ & $1.33 (5)$ & $\frac{4}{3}$ & $1.36 (2)$ & $\frac{11}{8}$ & $1.40 (2)$ & $\frac{17}{12}$ & $1.48 (2)$ & $\frac{3}{2}$ \\

${SC}$ & $0.75 (2)$ & $\frac{3}{4}$ & $0.55 (3)$ & $\frac{13}{24}$ & $0.35 (7)$ & $\frac{7}{20}$ & $0.03 (8)$ & $0$\\
${G}$ & $-0.90 (5)$ & $-\frac{11}{12}$ & $-0.71 (5)$ & - & $-0.63 (5)$ & - & $-0.59 (5)$ & -\\
\hline
\hline
\end{tabular}
\caption{Comparison of the numerical estimates (n) for the subset fractal
dimensions $D_S$ with the exact predictions (e) where available.
Uncertainties of the last decimal(s) for each $D_S$ are given in parenthesis.}
\label{numerics.tab}
\end{center}
\end{table}


\begin{figure}[!ht]
\includegraphics[width=8cm]{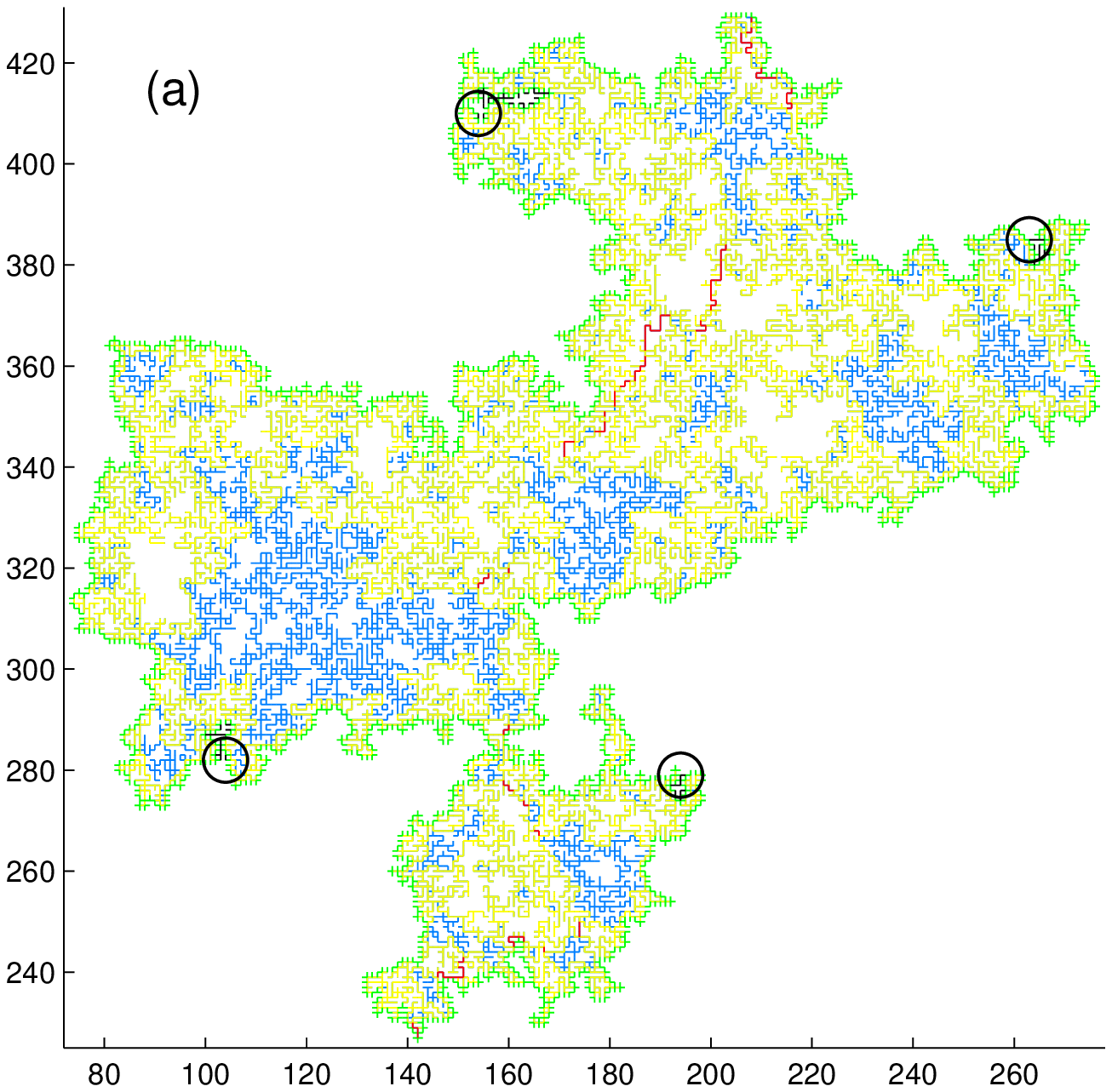}
\end{figure}

\begin{figure}[!ht]
\includegraphics[width=8cm]{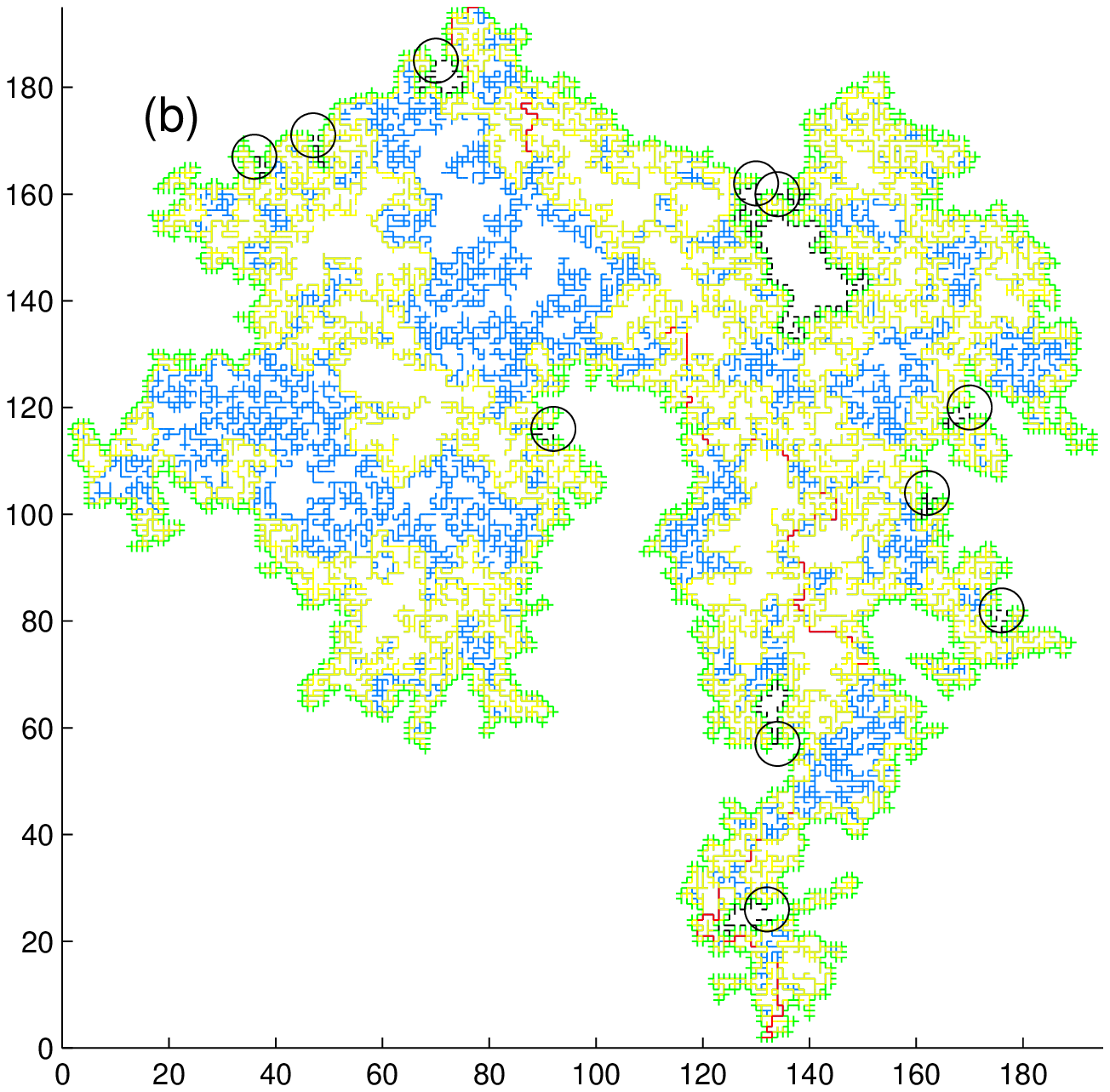}
\end{figure}

\begin{figure}[!ht]
\includegraphics[width=8cm]{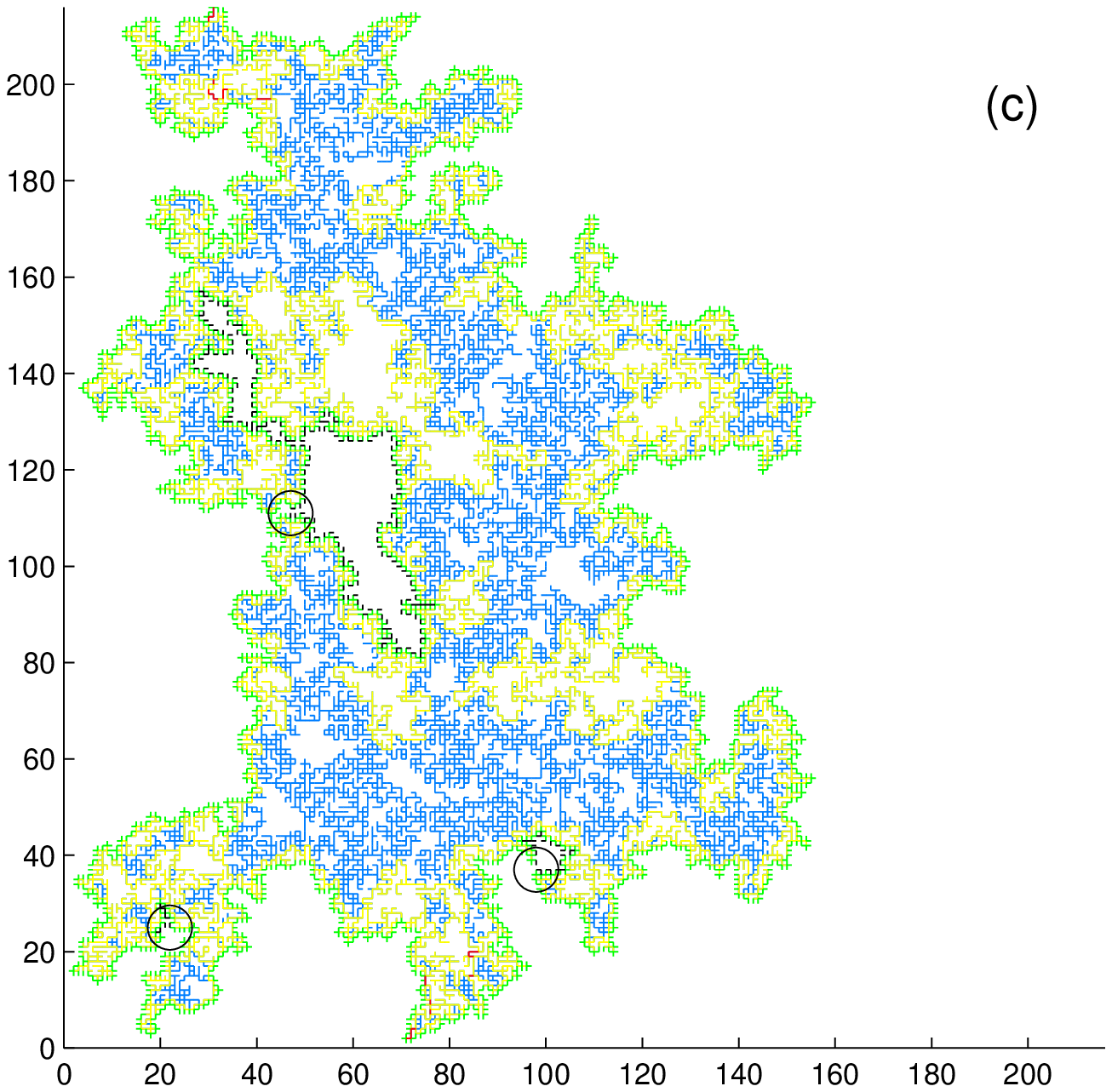}
\end{figure}

\begin{figure}[!ht]
\includegraphics[width=8cm]{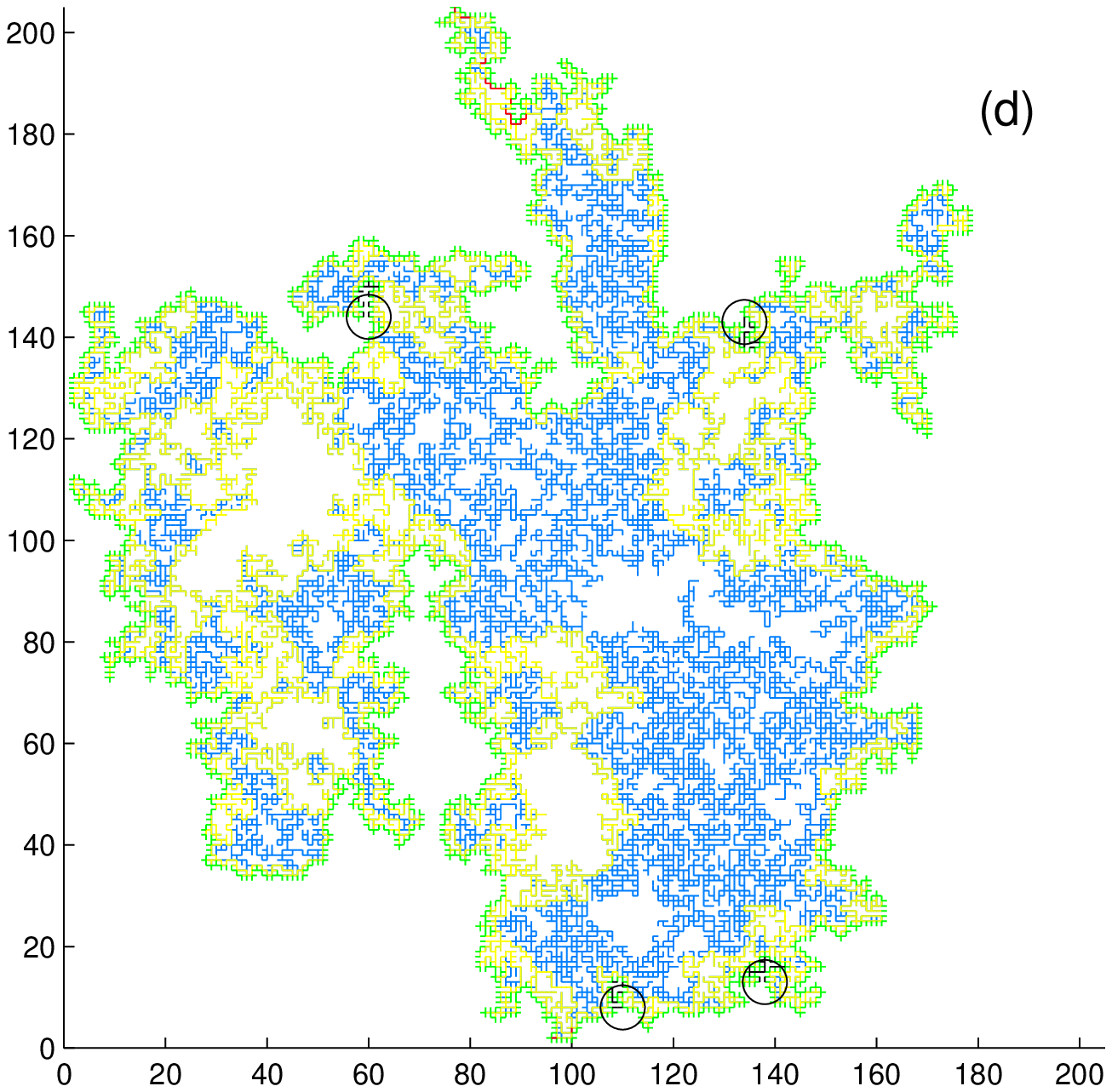}
\caption{Computer generated Potts clusters for (a) $q=1$, (b)
$q=2$, (c) $q=3$ and (d) $q=4$ state Potts models. Colors indicate
different subsets: SC bonds are shown in red, H bonds are shown in
yellow and the rest of the bonds contributing to M are shown in
blue. The EP bonds are colored green and the gates to fjords are
marked by black circles, while the fjord is shown with a black
line. For all the clusters, the total masses $M_M$ are in the
range $14400 - 17600$. Note the decrease of $D_H$ with $q$.}
\label{fig1}
\end{figure}

\begin{figure}[!ht]
\includegraphics[width=8cm]{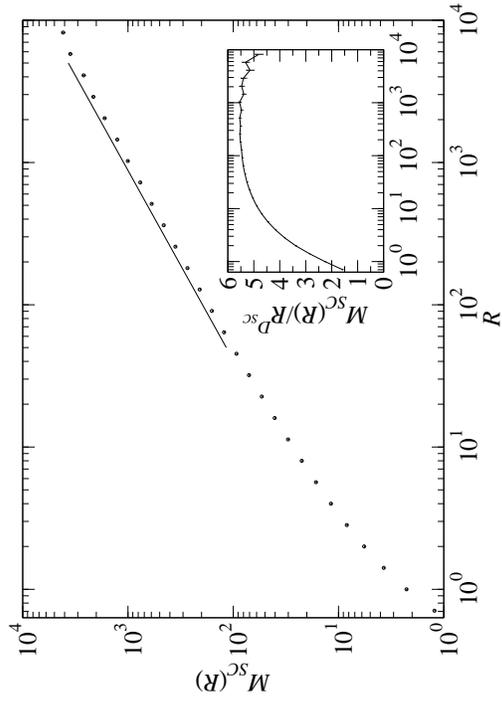}
\caption{Site percolation model. The number of the singly
connected bonds $M_{SC}(R)$ vs. the cluster linear size $R$. The
predicted slope $D_{SC}=3/4$ is indicated by the solid line. The
inset shows the scaled mass $M_{SC}(R)/R^{D_{SC}}$. Note the
saturation to the asymptotic scaling at $R \approx 300$.}
\label{fig2}
\end{figure}

\begin{figure}[!ht]
\includegraphics[width=8cm]{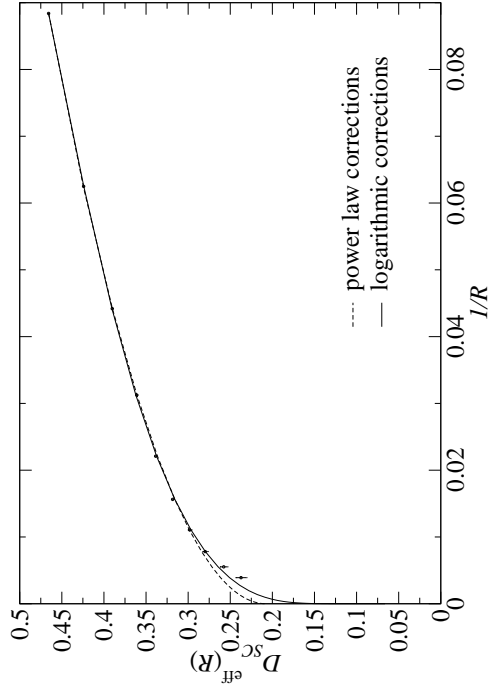}
\caption{Example of different fits in the $q=4$ Potts model to
$M_{SC}(R)$. The dashed line shows a fit with a single power law
correction ($\theta=1/2$) and the solid line shows the fit to the
logarithmic form of Eq. \eqref{logfit.eq}. Note the difference in
the extrapolation to the $R \rightarrow \infty$ where the fits
give $D_{SC} = 0.21 \pm 0.01$ with a single power law correction term
whereas  $D_{SC }= 0.03 \pm 0.08$ with the logarithmic form.}
\label{fig3}
\end{figure}

\begin{figure}[!ht]
\includegraphics[width=8cm]{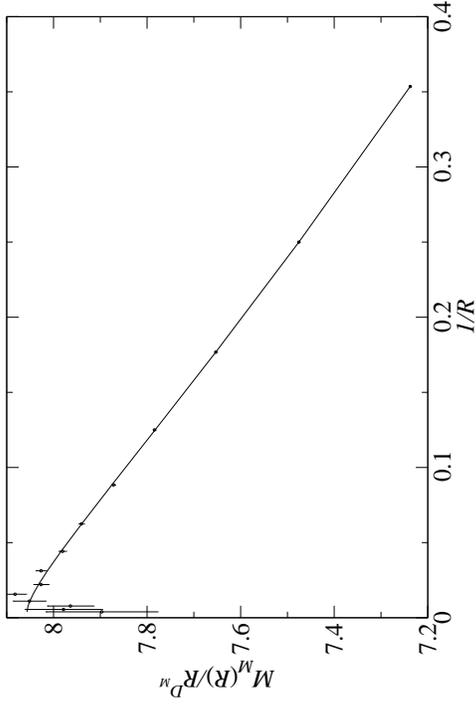}
\caption{Total mass of the cluster mass $M_M$ in the $3$-state Potts model.
Solid line is the nonlinear fit to the data. For this particular
fit $\chi^2=1.17$.}
\label{fig4}
\end{figure}

\begin{figure}[!ht]
\includegraphics[width=8cm]{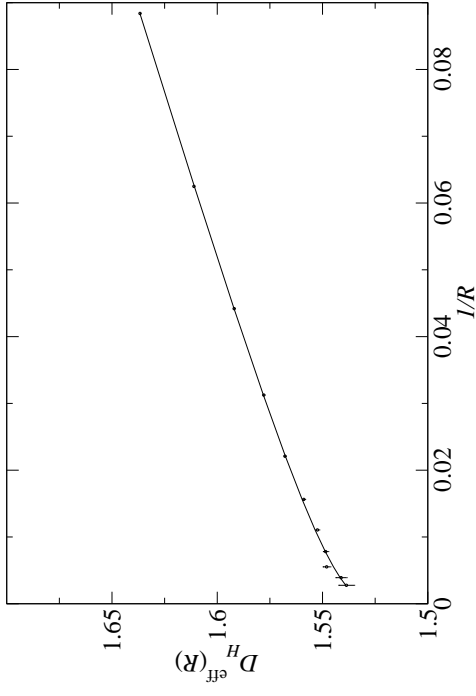}
\caption{Effective exponent $D_H^{\rm eff}(R)$ against $1/R$ in
the $q=4$ Potts model. For the fit indicated by the solid line we
find $\chi^2=1.11$.}
\label{fig5}
\end{figure}

\begin{figure}[!ht]
\includegraphics[width=8cm]{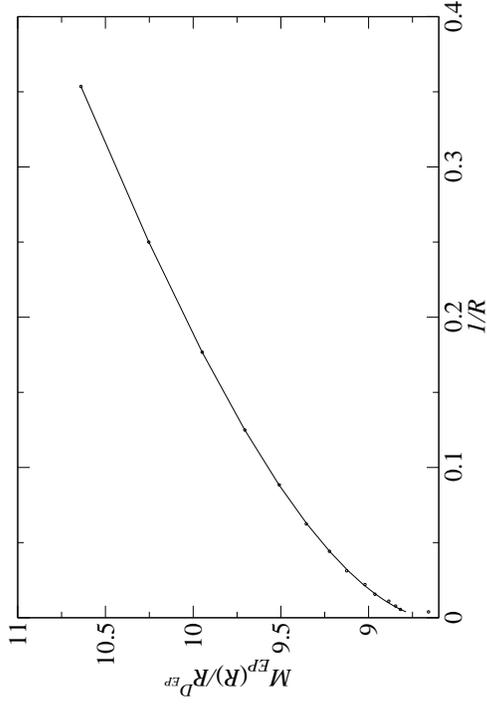}
\caption{Number of the external perimeter bonds, $M_{EP}$ versus
$1/R$ in the $q=2$ Potts model. Solid line indicates the fit to the data
($\chi^2=1.77$).}
\label{fig6}
\end{figure}

\begin{figure}[!ht]
\includegraphics[width=8cm]{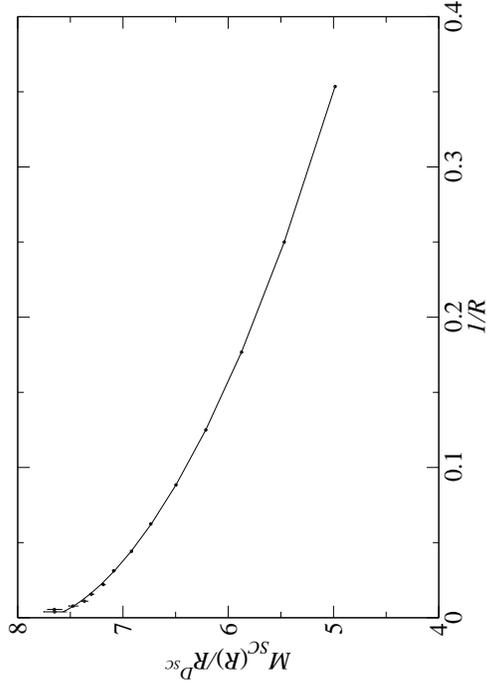}
\caption{Number of the singly connected bonds $M_{SC}$ against $1/R$
in the $q=2$ Potts model. Solid line is the nonlinear fit
for which $\chi^2=1.21$.}
\label{fig7}
\end{figure}

\begin{figure}[!ht]
\includegraphics[width=8cm]{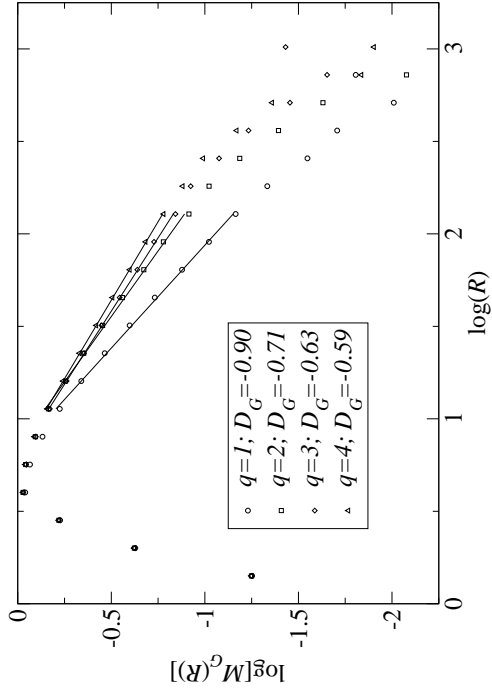}
\caption{Data for $M_G(R)$, the number of gates to fjords on
log-log scale. Different values of $q$ are represented by the
symbols shown in the legends. Straight lines indicate the fits to
the data; slopes give the exponents $D_G$.}
\label{fig8}
\end{figure}

\end{document}